\documentclass[final,5p,times,twocolumn]{elsarticle}

\usepackage{lineno}
\usepackage{amsmath, amssymb, amsfonts, latexsym}
\usepackage{subcaption}
\usepackage{graphicx}

\journal{Nuclear Instruments and Methods}

\begin{document}


\begin{frontmatter}


\title{Density Changes in Low Pressure Gas Targets for Electron Scattering Experiments}

\author[UNH]{S.~N.~Santiesteban}
\author[Kent]{S.~Alsalmi}
\author[JLab]{D.~Meekins}
\author[WM]{C.~Ayerbe Gayoso}
\author[TN]{J.~Bane}
\author[WM]{S.~Barcus}
\author[SM]{J.~Campbell}
\author[FI]{J.~Castellanos}
\author[MIT]{R.~Cruz-Torres}
\author[VT]{H.~Dai}
\author[Kent]{T.~Hague}
\author[OD]{F.~Hauenstein}
\author[JLab]{D.~W.~Higinbotham\corref{cor1}}
\author[argonne,CalTech]{R.~J.~Holt}
\author[SB]{T.~Kutz}
\author[UNH]{S.~Li}
\author[COL]{H.~Liu}
\author[JLab]{R.~E.~McClellan}
\author[Kent]{M.~Nycz}
\author[UVA]{ D.~Nguyen}
\author[hampton]{B.~Pandey}
\author[VT]{V.~Pandey}
\author[MIT]{A.~Schmidt}
\author[Kent]{T.~Su}
\author[argonne]{Z.~Ye}

\cortext[cor1]{Corresponding Author: doug@jlab.org}

\address[UNH]{University of New Hampshire, Durham, New Hampshire 03824, USA}
\address[Kent]{Kent State University, Kent, Ohio 44240, USA}
\address[JLab]{Jefferson Lab, Newport News, Virginia 23601 USA}
\address[TN]{The University of Tennessee, Knoxville, Tennessee 37996, USA}
\address[WM]{The College of William and Mary, Williamsburg, Virginia 23187, USA}
\address[SM]{Saint Mary's University, Halifax, Nova Scotia, Canada}
\address[FI]{Florida International University, Miami, Florida 33199 USA}
\address[MIT]{Massachusetts Institute of Technology, Cambridge, Massachusetts 02139, USA}
\address[VT]{Center for Neutrino Physics, Virginia Tech, Blacksburg, Virginia 24061, USA}
\address[OD]{Old Dominion University, Norfolk, Virginia 23529, USA}
\address[CalTech]{Kellogg Radiation Laboratory, California Institute of Technology, Pasadena California 91125 USA}
\address[argonne]{Physics Division, Argonne National Laboratory, Argonne, Illinois 60439, USA}
\address[COL]{Columbia University, New York, New York 10027, USA}
\address[SB]{Stony Brook University, Stony Brook, New York 11794, USA}
\address[UVA]{Department of Physics, University of Virginia, Charlottesville, Virginia 22904, USA}
\address[hampton]{Hampton University, Hampton, Virginia 23669, USA}

\begin{abstract}
A system of modular sealed gas target cells has been developed for use in electron scattering experiments 
at the Thomas Jefferson National Accelerator Facility (Jefferson Lab). This system was initially developed 
to complete the MARATHON experiment which required, among other species, tritium as a target material. 
Thus far, the cells have been loaded with the gas 
species $^{3}$H, $^{3}$He, $^{2}$H, $^{1}$H and $^{40}$Ar and operated in nominal beam currents of up to $22.5$~$\mu$A
in Jefferson Lab's Hall~A.  
While the gas density of the cells at the time of loading is known, 
the density of each gas varies uniquely when heated by the electron beam.  To extract experimental cross sections using 
these cells, density dependence on beam current of each target fluid must be determined.  
In this study, data from measurements 
with several beam currents within the range of $2.5$ to $22.5$~$\mu$A on each target fluid are presented. 
Additionally, expressions for the beam current dependent fluid density of each target are developed.
\end{abstract}

\begin{keyword}
target \sep
\sep tritium
\sep helium 
\sep deuterium
\sep hydrogen
\sep argon
\end{keyword}
\end{frontmatter}


\section{Introduction}

A modular gas cell target system was developed for use in Jefferson Lab's Hall~A for 
the MARATHON experiment E12-10-103~\cite{marathon}.
The design was specifically developed to safely contain and operate with gaseous tritium.  
The modular design allows gas cells filled with other species of gas 
to be installed in the system concurrently. The target was 
also adapted for experiments E12-11-112 ($x_{b}>1$)~\cite{E12-11-112}, 
E12-14-011 ($e,e'p$)~\cite{E12-14-011,Cruz-Torres:2019bqw}, 
E12-17-003 (Hypernuclear)~\cite{hypernuclear} and E12-14-009 (elastic)~\cite{E12-14-009}.  
MARATHON, together with these experiments, became known as the tritium group of experiments 
and were performed from December 2017 through November 2018.  Prior to the tritium group
of experiments operations, 
a target cell of this same design was filled with argon gas and used by
experiment E12-14-012 (Argon)~\cite{E12-14-012,Dai:2018gch} during Spring of 2017. 

While the performance of the target was an important consideration, the primary objective 
of the target system design and construction was to ensure safe operations with tritium 
gas under all conditions.   These conditions included target cell preparations, 
loading, storage, transportation, installation, removal, and beam operations. 
This was accomplished with a modular design, rigorous fabrication and testing, proper 
quality assurance and quality control, and multiple layers of containment/confinement.    
 
In addition to describing of the target, we present the beam current dependent density of the five gases used 
with the target system, $^{3}$H, $^{3}$He, $^{2}$H, $^{1}$H and $^{40}$Ar. 
The electron beam deposits energy in the cell end caps as well as in the target fluid.
This ionization energy, which is proportional 
to the beam current, heats the target fluid causing local changes in the density. To determine the magnitude of this 
effect, data were collected with the left high resolution spectrometer (LHRS) in Jefferson Lab Experimental Hall~A during 
February 2017 for the $^{40}$Ar target and December 2017 for the other targets. The beam energy for the study was $2.2$ GeV 
in all cases, The angle and momentum settings were $17.5 ^\circ $ and $1.79$ GeV for $^{40}$Ar, and $17.0 ^\circ $ and $1.99$ GeV 
for the other fluids.  Analysis shows that a simple quadratic polynomial function normalized to zero current 
provides an excellent descriptive fit function for all target fluids.

\section{Target System}

\begin{figure}[htbp]
\centering
\includegraphics[width=0.9\linewidth]{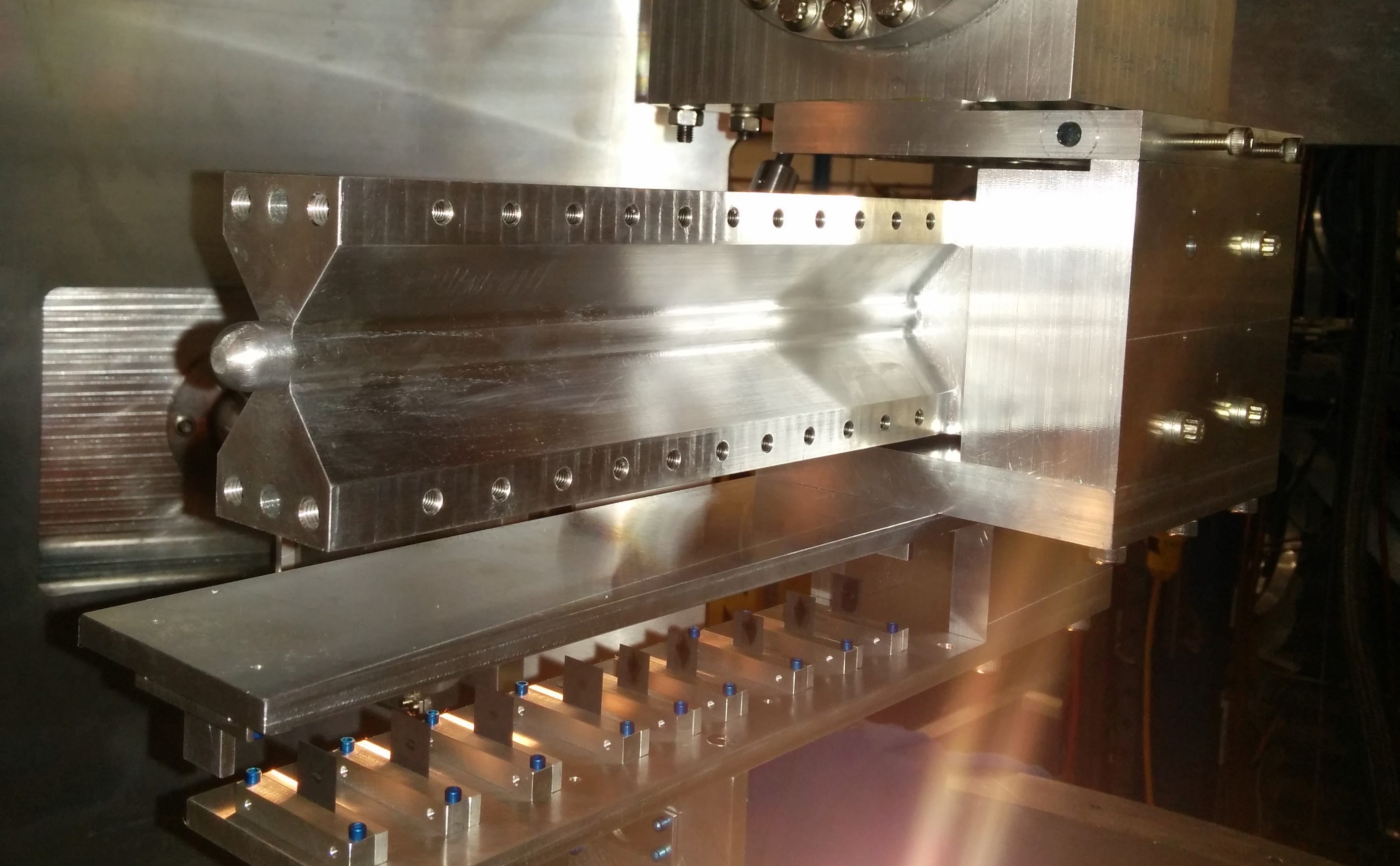}\\
\caption{A photo of the cell filled with $^{40}$Ar installed 
         on the standard Hall~A cryogenic target ladder.   Below
         the gas cell can be seen the carbon foil target which are used 
         to calibrate the reconstruction matrix of the spectrometers.}
\label{argon}
\end{figure}

\begin{figure}[htb]
\centering
\includegraphics[width=\linewidth]{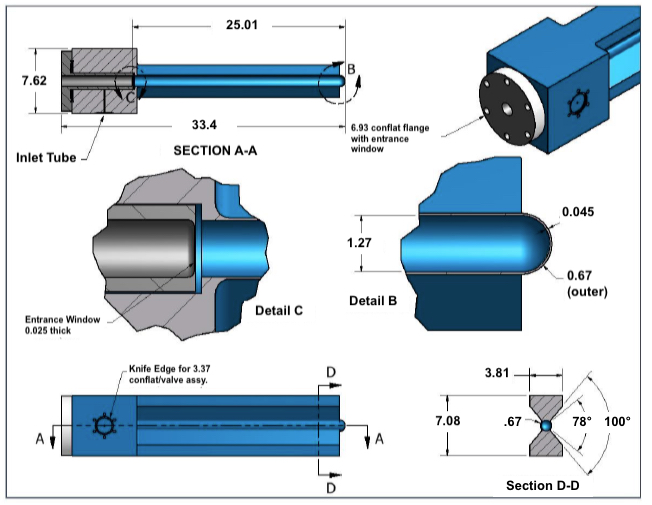}
\caption{Overview of the design of the gas target cells with the units in cm.}
\label{fig:celldesign}
\end{figure}

The modular design allows for multiple cell configurations. It also enables individual cells to be installed in special configurations 
of the standard Hall~A cryogenic target such as the $^{40}$Ar target (see Fig.~\ref{argon}). 
Another feature is that 
it allows cells to be filled at off-site locations.  The tritium cell was filled at Savannah River 
Site (SRS) by Savannah River Tritium Enterprises (SRTE), with 0.1~grams
of tritium gas to a room temperature absolute  pressure of 1.38~MPa. It was shipped 
in a special purpose transport container called the bulk tritium shipping package (BTSP). Including the cell this system provided 
continuous triple layer confinement throughout the shipping and handling
process. This design also allowed the tritium cell to be placed  in a 
storage container in Hall~A while normal Hall installation activities were
completed. The tritium cell was installed after all other preparatory tasks were completed.     
The modular sealed gas cell represents a departure from previous designs~\cite{Beck:1989bi}. 
Fig.~\ref{fig:celldesign} shows the design of the gas cell design.  
This design is similar to the design proposed in Ref.~\cite{Brajuskovic:2013ymh} with engineering
details for the construction and loading of these cell in Ref.~\cite{engreport}.
.

\begin{figure}[htbp]
\centering
\includegraphics[width=0.8\linewidth]{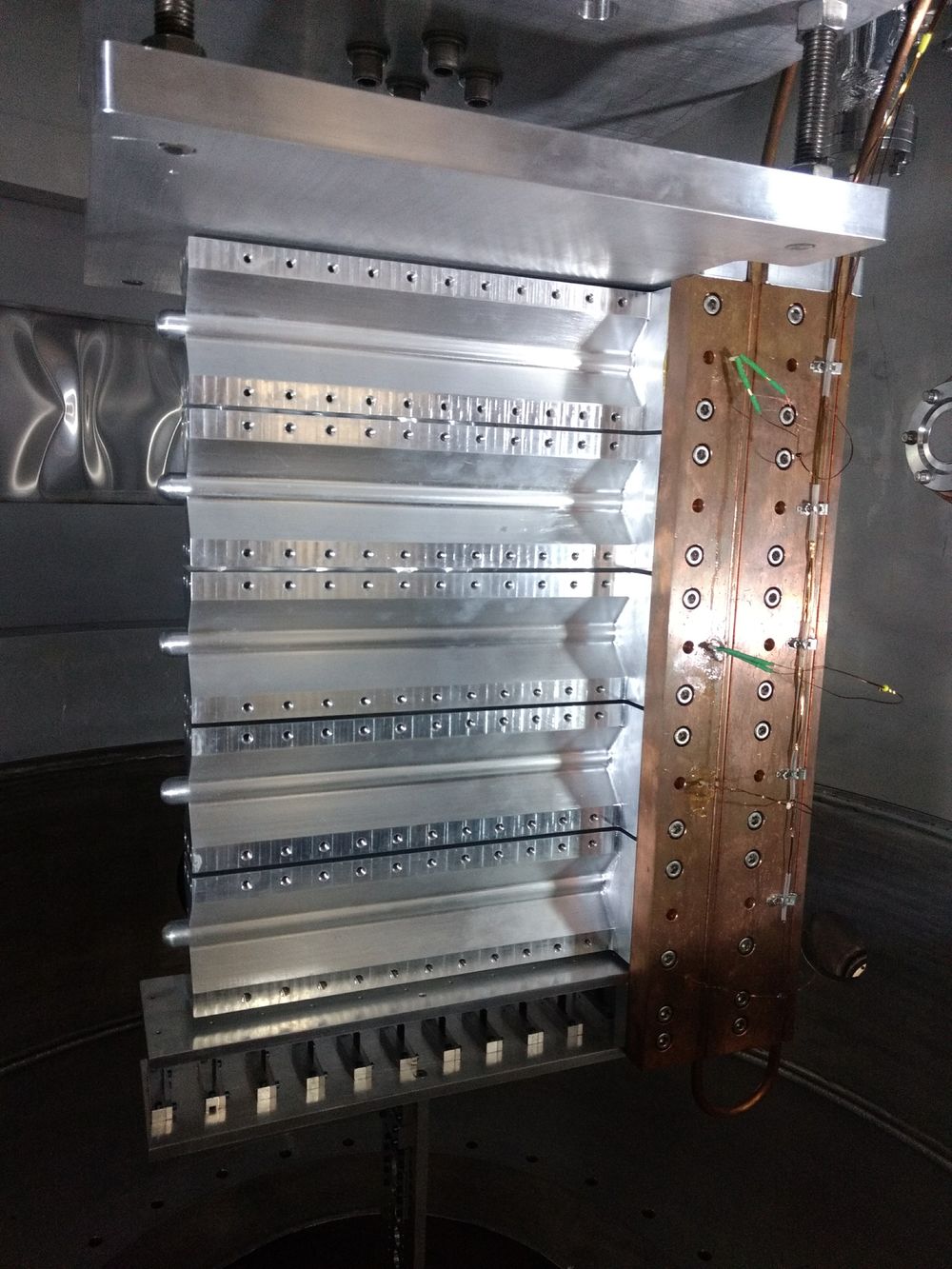}\\
\caption{Ladder assembly showing the five cells that were installed for the MARATHON experiment, $^{3}$H, $^{2}$H, $^{1}$H, $^{3}$He and empty cell from top to bottom, 
        as assembled during Fall 2017 to Spring 2018 run period.}
\label{ladder}
\end{figure}

The configuration of the target system for the tritium experiments is shown in Fig.~\ref{ladder}. 
In this configuration, there are (from top to bottom) 
four cells loaded with $^{3}$H, $^{2}$H, $^{1}$H, and $^{3}$He as well as a fifth empty cell which was used for background 
measurements.  The cells are contained in a scattering chamber which is under vacuum. 
The scattering chamber vacuum is isolated from the upstream beam line vacuum by a 0.2~mm 
thick beryllium window. This window is roughly 30~cm upstream of the target center 
and is mounted on a reentrant tube that also contains a 15~cm long tungsten collimator 
with an inner diameter of 12.7~mm.  The scattering chamber vacuum, with a pumping system 
directed to an exhaust stack, provided a second layer of tritium confinement. An 
exhaust system, (together with strict access controls) capable of maintaining a slight 
negative pressure in the experimental Hall ensured that the Hall boundary 
was a third layer of confinement.

Each cell is machined from ASTM B209 aluminum 7075-T651 plate. 
Each target cell has a cylindrical fluid space with a length of 25~cm and a 
diameter of 12.7~mm. The total volume of the cell (including the non-active region) 
is $33.4 \pm 0.2$~cm$^{3}$. The thickness of the nearly flat entrance window and hemispherical 
exit window is nominally 0.25~mm.  The parameters at the time of loading for each cell 
are summarized in Table~\ref{tab:fill_tar}. 

\begin{figure}[htb]
\centering
\includegraphics[width=0.85\linewidth]{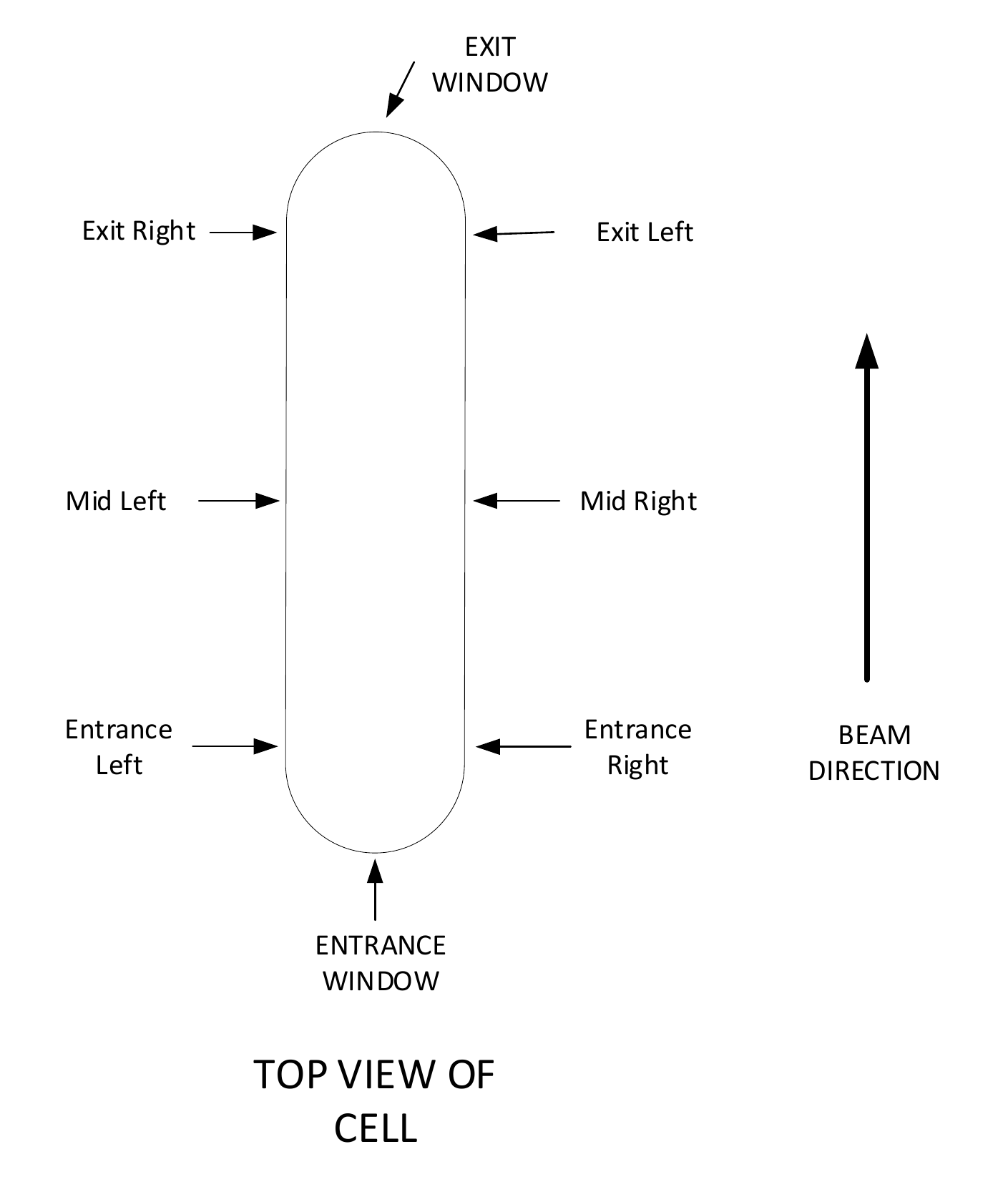}\\
\caption{Measurement locations of the cells represented schematically.}
\label{fig:cellconfig}
\end{figure}

Due to machining tolerances, the wall thickness of each cell varies slightly over its length. 
Thickness measurements were performed for each cell at several 
locations as schematically represented in Fig.~\ref{fig:cellconfig} and summarized 
in Table~\ref{tab:cell}.
These measurements were performed with a Magna Mike 8600 Hall effect thickness gauge 
which provides a relative uncertainty of 0.001~mm and an absolute uncertainty of 0.007~mm. 
The error shown in the Table~\ref{tab:cell} indicates the standard deviation of multiple measurements 
in a 2~mm radius for a given location.
The $^{40}$Ar cell, installed in February 2017, was later 
evacuated and installed as the empty cell for the tritium group 
of experiments, so  Table~\ref{tab:cell} shows the $^{40}$Ar and the empty cell in a single column. 

Once installed in the Hall A scattering chamber,  the target cells were cooled to 40~K 
with the temperature maintained using a 15~K helium supply and a controlled heater. 
This cooling was required to removed the modest amount of heat generated 
by the electron beam passing through the target fluid, cell entrance and cell exit, which, in total, 
was about 15~W.  To ensure cell integrity, the maximum beam current permitted on any of the cells 
was $22.5$ $\mu$A~\cite{engreport}.  The heat generated by the tritium decay is very small, 
about 50~mW.

\begin{table}[htb]
\centering
\begin{tabular}{cccc}
	\hline 
	Target       & Fill Pressure & Fill Temp    & Thickness \\
	     		 &	(kPa)		 &	(K) 	    & (mg/cm$^2$) \\
	\hline 
	$^{40}$Ar	 & $3447$ 		 & $291.0$	    &  $1455\pm9$ \\ 
	
	$^{3}$H $1^{st}$ 	 & $1400$		 & $296.3$	    &  $85.1\pm 0.8$ \\ 
	
	$^{3}$H $2^{nd}$	 & $1393$		 & $293.8$		&  $84.8\pm0.8$ \\
	
	$^{3}$He	 & $1772$		 & $294.3$	    &  $53.4\pm0.6$ \\ 
	\
	$^{2}$H 	 & $3549$		 & $296.1$	    &  $142.2\pm0.8$ \\ 
	
	$^{1}$H 	 & $3549$		 & $297.4$	    &  $70.8\pm0.4$ \\ 
	\hline 
\end{tabular}
\caption{The target thickness in gm/cm$^2$ for each of the gas cells based on the fill pressures and temperatures. 
Temperatures have an uncertainty of 0.1~K.}
\label{tab:fill_tar}
\end{table}

\begin{table*}[htb]
\centering
\begin{tabular}{cccccc}
\hline
Location        & \begin{tabular}[c]{@{}c@{}} $^{40}$Ar/Empty Cell\\ Thickness (mm)\end{tabular} & \begin{tabular}[c]{@{}c@{}}$^{3}$H Cell\\ Thickness (mm)\end{tabular} & \begin{tabular}[c]{@{}c@{}}$^{1}$H Cell\\ Thickness (mm)\end{tabular} & \begin{tabular}[c]{@{}c@{}}$^{2}$H Cell\\ Thickness (mm)\end{tabular} & \begin{tabular}[c]{@{}c@{}}$^{3}$He Cell\\ Thickness (mm)\end{tabular} \\ \hline
Entrance               &      $0.254 \pm 0.005$                                                  & $0.253 \pm 0.004$                                                     & $0.311 \pm 0.001$                                                     & $0.215 \pm 0.004$                                                     & $0.203 \pm 0.007$                                                      \\ 
Exit                    &    $0.279 \pm 0.005$                                                  & $0.343 \pm 0.047$                                                     & $0.330 \pm 0.063$                                                     & $0.294 \pm 0.056$                                                     & $0.328 \pm 0.041$                                                      \\ 
Exit left              &      $0.406 \pm 0.005$                                                 & $0.379 \pm 0.007$                                                     & $0.240 \pm 0.019$                                                     & $0.422 \pm 0.003$                                                     & $0.438 \pm 0.010$                                                      \\ 
Exit right              &       $0.421 \pm 0.005$                                                 & $0.406 \pm 0.004$                                                     & $0.519 \pm 0.009$                                                     & $0.361 \pm 0.013$                                                     & $0.385 \pm 0.016$                                                      \\ 
Mid left             &        $0.457 \pm 0.005$                                                  & $0.435 \pm 0.001$                                                     & $0.374 \pm 0.004$                                                     & $0.447 \pm 0.009$                                                     & $0.487 \pm 0.060$                                                      \\ 
Mid right             &        $0.432 \pm 0.005$                                                  & $0.447 \pm 0.004$                                                     & $0.503 \pm 0.005$                                                     & $0.371 \pm 0.012$                                                     & $0.478 \pm 0.007$                                                    \\ 
Entrance left      &          $0.508 \pm 0.005$                                                  & $0.473 \pm 0.003$                                                     & $0.456 \pm 0.010$                                                     & $0.442 \pm 0.005$                                                     & $0.504 \pm 0.003$                                                      \\ 
Entrance right     &		 $0.424 \pm 0.005$                             & $0.425 \pm 0.003$                                & 
$0.457 \pm 0.006$                                & 
$0.332 \pm 0.011$                                & 
$0.477 \pm 0.011$                                 \\ \hline
\end{tabular}
\caption{Cell wall thickness measurements for different cells as measured by a Hall effect thickness gauge.}
\label{tab:cell}
\end{table*}

\section{Hall~A Spectrometers}

The data were acquired with the left high resolution spectrometer (LHRS). 
For a detailed description of the LHRS see Ref.~\cite{Alcorn:2004sb}. 
The basic components of the LHRS 
are a normal conducting quadrupole (Q1), a superconducting quadrupole (Q2), a superconducting dipole (D), 
and a superconducting 
quadrupole (Q3) in a Q-Q-D-Q configuration. The quadrupoles focus scattered charged particles while the 
dipole bends these particles, 
within a given momentum range, to the detectors. After passing through the spectrometer magnets, 
the scattered particles pass through 
two vertical drift chambers (VDCs) that provide tracking information~\cite{Fissum:2001st}. 
Two layers of scintillator hodoscopes, s0 and s2, are on either side of a gas 
Cherenkov detector filled with CO$_2$~\cite{Iodice:1998ft}.  The hodoscopes provide trigger and time of 
flight for the detected particles.  
The Cherenkov provides identification of electrons with approximately $99\%$ efficiency 
and reject $\pi^{-}$ below a momentum of 4.8~GeV/c.  
The last element in the detector stack is the shower calorimeter. Electrons passing through the calorimeter lead glass blocks 
induce a cascade of pair production and bremsstrahlung radiation from which their energy 
can be determined~\cite{Alcorn:2004sb}.

\section{Beam Current Monitor}
\label{BCM}

The beam current monitor (BCM) is a system of three independent devices and a current source~\cite{Denard:2001zg}. 
This is a dedicated system in Hall~A and while independent of the target effects, this system
is the dominant source of systematic uncertainty in the current dependent density studies presented
herein.  The BCM system consists of a toroidal sensor (Unser)~\cite{Unser:1991dr}, located between 
upstream and downstream RF cavities, and a data-acquisition system.  A current source, which is connected to a wire which 
passes through the Unser, is used to calibrate the Unser immediately prior to each use of the device and
the Unser is then used to calibrate the BCMs with the electron beam. 

The Unser monitor is composed of two identical toroidal coils driven in opposite directions by an external source.  
The DC component of the current flowing through the toroid sensor is detected by a magnetic modulator. The 
beam current passing through the cores produces a flux imbalance, which generates an output signal proportional to the 
even harmonics of the frequency of excitation, In the absence of a DC current, the sum 
of the signals is zero~\cite{Denard:2001zg}. 

\begin{figure}[htb]
\centering
\includegraphics[width=0.9\linewidth]{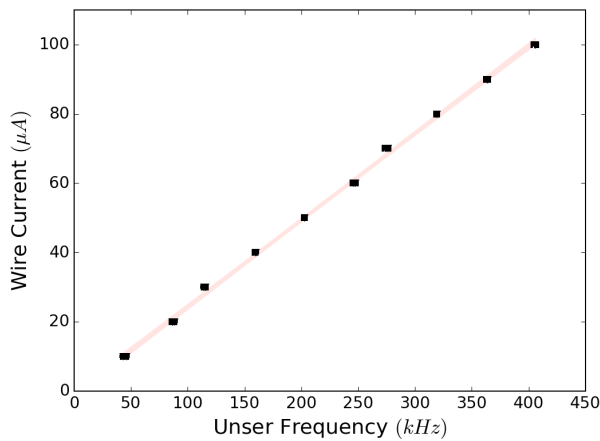}
\caption{Wire Unser calibration. The band represents the $95\%$ confidence level of the linear fit.}
\label{fig:unser_cal}
\end{figure}

The temperature controlled Unser has a sensitivity to beam current of about 4~$mV/\mu$A and has a
DC offset subtracted stability within $0.1\%$~\cite{Denard:2001zg}.  The systems DC offset does slowly 
drift, necessitating
the current calibration to be done immediately prior to using it for an absolute current calibration
of the RF cavities.  Once calibrated, the RF cavities are used to continuously monitor the beam current. 
The calibrations are checked periodically throughout the course of an experiment. 
To put the signals from the Unser and RF cavities into the scalers of Hall's fast data acquisition system,
a voltage to frequency (V/F) converter is used along with a discriminator. 
Figure \ref{fig:unser_cal} shows the Unser calibration with a known DC current source, 
the response of the system is found to be ($249.7 \pm 9.6) \times 10^{-6}$~$\mu$A/Hz. 

\begin{figure}[htb]
\centering
\includegraphics[width=0.9\linewidth]{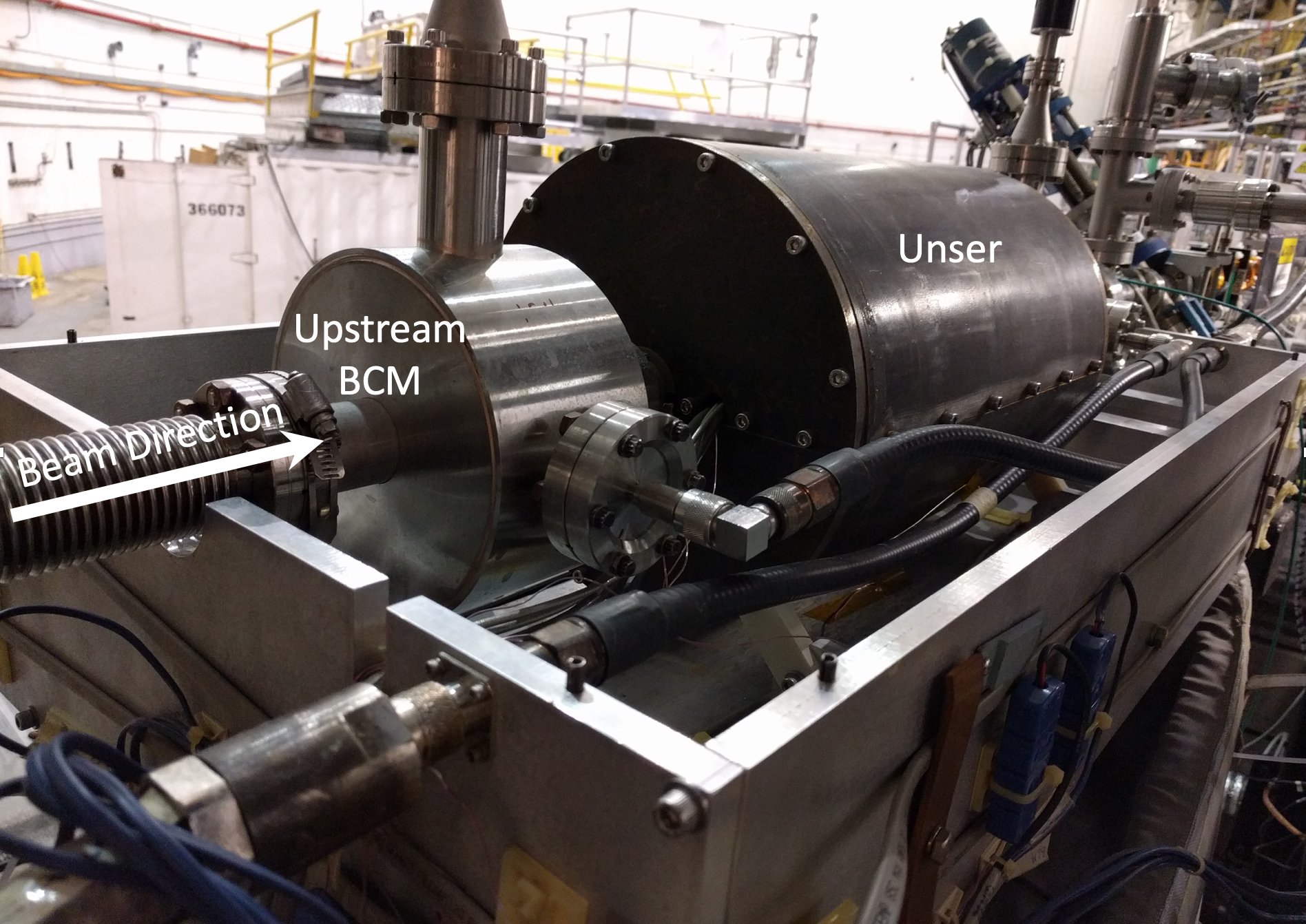}
\caption{Shown is the upstream beam current monitor (BCM) and the Unser cavity as installed in the Hall~A beamline.   
The thermal insulating cover, that keeps the systems at a stable temperature, has removed for the photo.}
\label{UnserPhoto}
\end{figure}

\begin{figure}[htb]
\centering
\includegraphics[width=\linewidth]{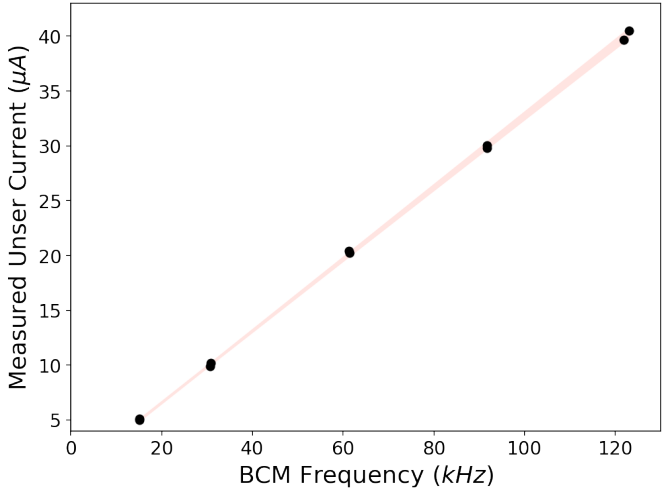}
\caption{BCM calibration data shown with the $95\%$ confidence level from a linear regression.}
\label{fig:dnew_cal}
\end{figure}

The beam current monitors (BCM) are 1497~MHz resonant cavities located immediately 
before and after the Unser and are 
used to continuously monitor the beam currents in Hall~A (see Fig.~\ref{UnserPhoto}).
The cavities are composed of loop antennas located where the magnetic field is maximum. 
When the beam passes through, the output RF signal is proportional to the current~\cite{Denard:2001zg}. 
As consequence, the BCM response is linear with respect to the current. Like the Unser, the signals 
from the RF cavities are filtered by a V/F converter. Several values of beam current (measured by the 
calibrated Unser) are used to the determine the linear dependence of the BCM as shown in 
Fig.~\ref{fig:dnew_cal}.  In general, the beam current can be then calculated using

\begin{equation}
I = g_{\rm{BCM}}\cdot f+O.
\label{eq:current_calc}
\end{equation}

\noindent where $g_{\rm{BCM}}$ and $O$ are the fit parameters, which correspond 
to ($326.4 \pm 1.4) \times 10^{-6}$~$\mu$A/Hz 
and $0.1 \pm 0.09$  $\mu A$, respectively. Finally, for any given beam induced frequency $f$, the 
current $I$ is given by Eq.~\ref{eq:current_calc}.  Unfortunately the BCM system becomes much 
less accurate for beam currents below $\sim 5$~$\mu$A.   

\section{Method Overview}

The density of the target is well known when loaded but experience and simulations have shown that the beam current will 
decrease the local density of the target fluid in the beam path.  The magnitude of this effect depends on the 
beam current and target fluid species and must be quantified to accurately determine cross sections, ratios and 
other comparisons of data collected with the multiple gas cells~\cite{celldes}. 
It was shown (with the exception of the argon cell) that the target density reaches equilibrium 
within a few seconds from when the 
electron beam first impinges on the cell and the density was constant with stable beam current. The purpose of 
these measurements and analysis is to develop
a calibration of the target density as a function of beam current for each gas species.

In order to extract the current dependent density correction, the LHRS is used to measure the
event rate for several beam currents. 
The normalized yield is determined by applications of corrections to the raw event rate. 
These corrections include: integrated charge during the measurement, particle identification, 
acceptance cuts, detector efficiencies, and live times.  The normalized yield $Y_{\rm{norm}}$ is then given by

\begin{equation}
Y_{\rm{norm}} = \frac{PS \cdot N}{ Q \cdot \epsilon \cdot LT },
\label{eq:yield}
\end{equation}

\noindent where $N$ is the number of good electrons, $PS$ is the prescale factor of the DAQ system, 
$Q$ is the integrated charge, and $\epsilon $ is the combined efficiency of the detectors, triggers 
and events selection cuts and $LT$ is the live-time. 
Each one of these parameters is explained in detail in the following sections.

\subsection{Event Selection}
To improve counting efficiency and maximize live time, a compound trigger was used. This trigger required both scintillator planes 
and the Cherenkov detector to have signals above threshold in order to exclude $\pi^{-}$ events.  To extract a good 
electron sample, several cuts were applied to the data. These cuts can be summarized in two groups: acceptance cuts, which 
assure that the events are selected within an acceptable spectrometer phase space, and 
tracking/particle identification (PID) cuts, which focus on the selection of electrons scattered from the target fluid.
These selection cuts are:

\begin{itemize}
\item[i.] Momentum and angular acceptance cuts: Specifically, the ranges used to determine $Y_{\rm{norm}}$ are $|\delta p/p| < 4.5\%$,  
$|\theta-\theta_{0}| < 30$~mrad and $|\phi| < 25$~mrad.

\item[ii.] Target length cut: This cut excluded events reconstructed back to the target windows 
and reduced background by limiting the effective target length $|y_{tar}|<8$~cm 

\item[iii.] Only events with a single track in the VDC were kept.

\item[iv.] A particle ID cut was applied to the Cherenkov ADC sum

\item[v.] A particle ID cut was applied to the shower calorimeter

\end{itemize}

\noindent With these particle selection cuts, we found that the results remained stable within 1$\%$ run to run.
The cut for the reaction vertex was chosen such that the contamination from the aluminum end-caps was 
smaller than 2\%.  The systematic effects of the aluminum background events were studied for all the 
targets are included in the systematic uncertainties.

\subsection{Estimation of Efficiencies} 
 
A number of efficiencies were applied to the data to produce $Y_{\rm{norm}}$.  For simplicity, in this analysis 
only electron events with one track in the VDC were selected.  
The ratio between the total number of electron events with one track and the total number of triggered 
electrons (including multi-track and non-track particles) defines the VDC efficiency. 

The trigger efficiency 
was calculated using another trigger type, where only both scintillators were required to record the events. 
In this sense, the difference between the main trigger and the efficiency trigger is the Cherenkov detector.  
The ratio between the events recorded with the main and the efficiency trigger corresponds to the trigger efficiency. 

The Cherenkov efficiency was calculated by selecting a sample of electrons detected in the calorimeter 
and determining the number of events that also were detected in the Cherenkov detector. The calorimeter efficiency 
was measured by selecting a sample of electrons in the Cherenkov detector and counting the number of 
these electrons that also fired the calorimeter.

For these measurements, the trigger, Cherenkov, and calorimeters efficiencies were $>99\%$.   The tracking
efficiency was dependent on the absolute rate in the LHRS and varied for 97 to 99\%.

\subsection{Live-Time Calculation } 

The live-time is related with the limitation of the speed of data acquisition system (DAQ) to record events. 
It depends on the electronics, computers and trigger rate and is calculated using the ratio of the number 
of events recorded over the total number of events seen by the trigger. 
Typical values for the live-time ranged 93-97\% depending the trigger rate in the left LHRS as well as
the DAQ prescale setting.

\subsection{ Total Charge}

The beam is not completely stable throughout the run; it may trip off or fluctuate over time.
Therefore, we obtained the calibration data when the beam was mostly stable, and only runs
where the average current is within a window of $\pm 2$~$\mu$A of the requested current are used. 
The charge is calculated by integrating the current over time using the BCM calibration result (see Section~\ref{BCM}).

\section{Solid Target Check}

The aim of the analysis is to measure the density change when the beam is on the gas targets using 
the yield analysis. In order to test the method, the same analysis is applied to a solid target.  
The $^{40}$Ar experiment used a carbon foil while for the Tritium experiments used an aluminum target. 
Unlike the fluid targets, the solid target density is not measurably affected by the beam current.

\begin{figure}[htbp]
\centering
\includegraphics[width=\linewidth]{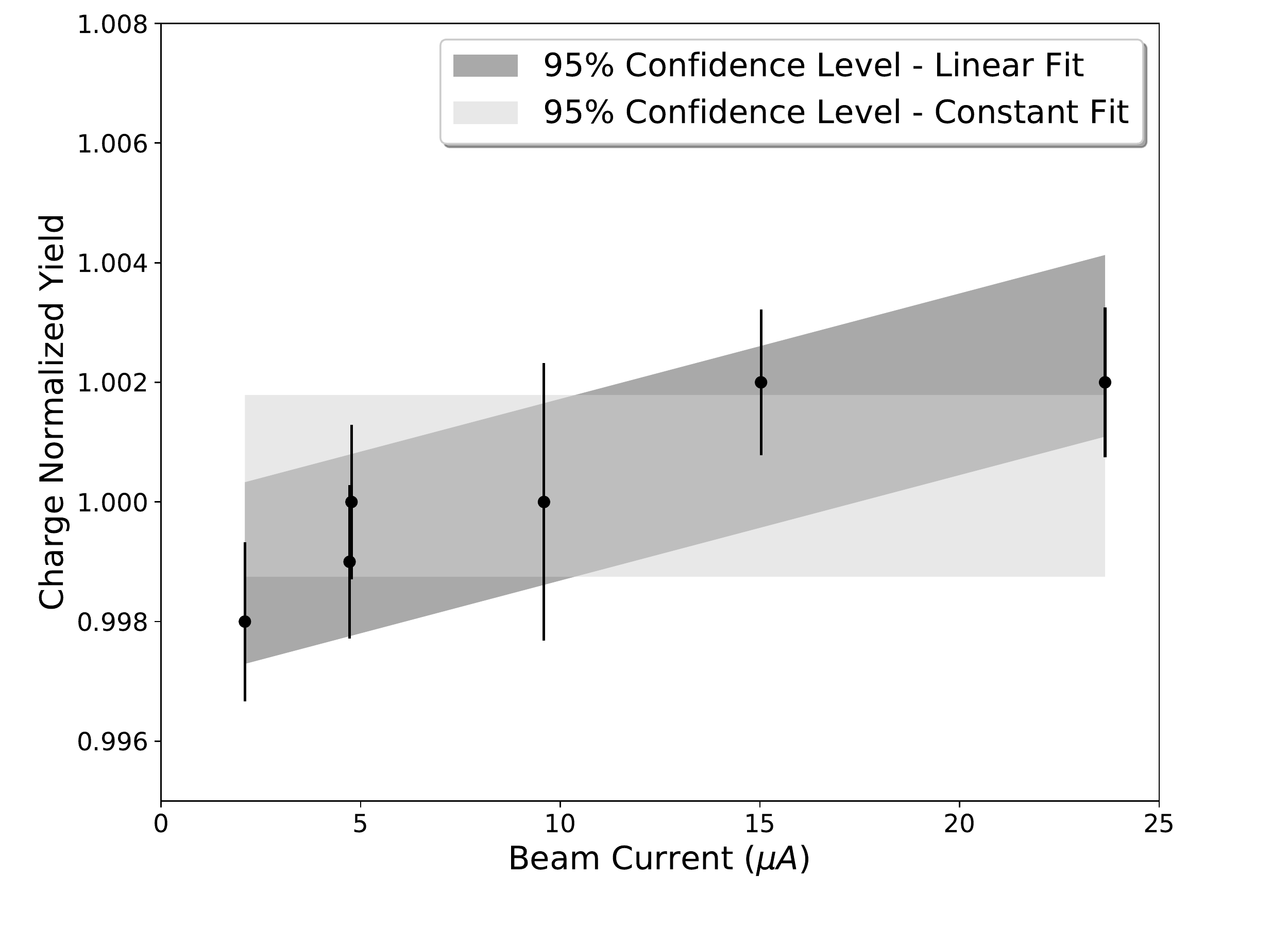}
\caption{Shown is the normalized yield vs. beam current the aluminum solid target used during the tritium group 
         experiments.   Shown are the 95\% confidence level bands for a constant and for a linear regression of the data.
         These results are consistent with are ability to determine density changes with
         our experimental setup to approximately 0.5\% at the 95\% confidence level.}
\label{fig:solid}
\end{figure}

Figure~\ref{fig:solid} shows $Y_{\rm{norm}}$ for the solid aluminum target which 
was calculated using Eq.~\ref{eq:yield} for 
different beam currents. It was normalized with respect to the lowest current yield value. 
The plot shows that  $Y_{\rm{norm}}$ 
did not change to within about 0.5$\%$ at the 95\% confidence level 
which is well within the uncertainty of the measurement. 

\section{Background Contamination}

The aluminum windows of the target cell contribute a background to the measured raw yield for each of the gas targets. 
To measure this background (henceforth referred to as contamination) in the case of the $^{40}$Ar experiment, 
a dummy target with aluminum foils with total thickness matching the radiation length of the argon filled cell was used.
In the case of the tritium experiments, an empty cell (or dummy cell)  was used. The normalized yields from 
these targets were then subtracted from the applicable $Y_{\rm{norm}}$. To check the current dependence of 
this subtraction, a comparison between the background at low and high current was measured for the dummy/empty targets. The charge yield given by Eq.~\ref{eq:current_calc} was binned in $y_{tar}$ segments along the target length, and the ratio of the events at high current to low current was determined. The ratio was found to be 1.006, which indicates that the background subtraction is independent of current, as expected. 

\begin{figure}[htb]
 \centering
 \includegraphics[width=0.9\linewidth]{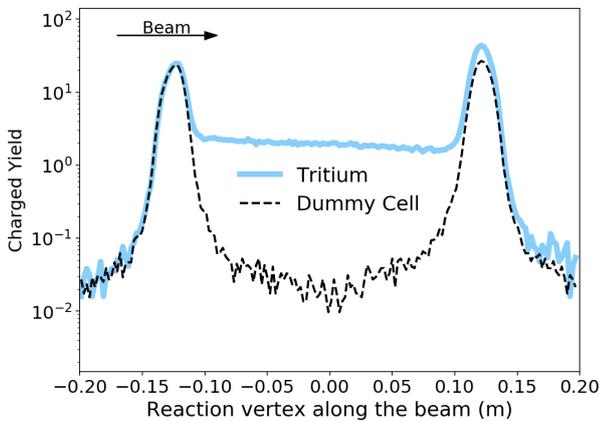}`
  \caption{Background contamination spectrum of the dummy target compared with that of tritium at 2.5~$\mu$A.  Both spectra are normalized.}
  \label{fig:bk_empty}
\end{figure}

Figure~\ref{fig:bk_empty} shows the spectra of the charge normalized yield for the empty (or dummy) cell 
and the tritium gas, for a beam current of $2.5$ $\mu A$. To optimize the signal to background ratio, 
events contributing to the $Y_{\rm{norm}}$ were selected from a symmetric region of $\pm 8$ $cm$ 
about the center of the target. Therefore, the contamination fraction is the ratio of $Y_{\rm{norm}}$ 
for the empty cell to $Y_{\rm{norm}}$ for the gas cell of interest. Table \ref{tab:contamination_al} 
summarizes the percentage of background contamination found in the gas targets for each beam current 
used in the study. 
 
\begin{table}[htb]
\begin{tabular}{ccccc|cc}
\hline
\textbf{\begin{tabular}[c]{@{}c@{}}Current \\ $(\mu A)$\end{tabular}} & \textbf{\begin{tabular}[c]{@{}c@{}}$^{3}$H\\ (\%)\end{tabular}} & \textbf{\begin{tabular}[c]{@{}c@{}}$^{3}$He\\ (\%)\end{tabular}} & \textbf{\begin{tabular}[c]{@{}c@{}}$^{2}$H\\ (\%)\end{tabular}} & \textbf{\begin{tabular}[c]{@{}c@{}}$^{1}$H\\ (\%)\end{tabular}} & \textbf{\begin{tabular}[c]{@{}c@{}}Current\\ $(\mu A)$\end{tabular}} & \multicolumn{1}{l}{\textbf{\begin{tabular}[c]{@{}l@{}}Argon\\ (\%)\end{tabular}}} \\ \hline
2.5                                                              & 1.7                                                             & 1.6                                                              & 0.7                                                             & 1.1                                                             & 2.5                                                             & 0.3                                                                                \\ 
5                                                                & 1.7                                                             & 1.6                                                              & 0.7                                                             & 1.2                                                             & 4.5                                                             & 0.3                                                                                \\ 
10                                                               & 1.7                                                             & 1.7                                                              & 0.8                                                             & 1.2                                                             & 8                                                               & 0.3                                                                                \\ 
15                                                               & 1.8                                                             & 1.8                                                              & 0.8                                                             & 1.3                                                             & 12                                                              & 0.3                                                                                \\ 
22.5                                                             & 1.8                                                             & 1.8                                                              & 0.8                                                             & 1.3                                                             & 15                                                              & 0.3                                                                                \\ 

 & & & & 
                                                                                                                                                                                                                                                                                                              & 18                                                              & 0.3                                                                                \\ \hline
\end{tabular}
\caption{Aluminum window contamination in a $\pm 8$ $cm$ range with respect to the center of the target at each nominal current. Note that these currents were not the same for both experiments.}
\label{tab:contamination_al}
\end{table}

\section{Gas Target Results}

\begin{figure}[htb]
	\centering
	\includegraphics[width=0.9\linewidth]{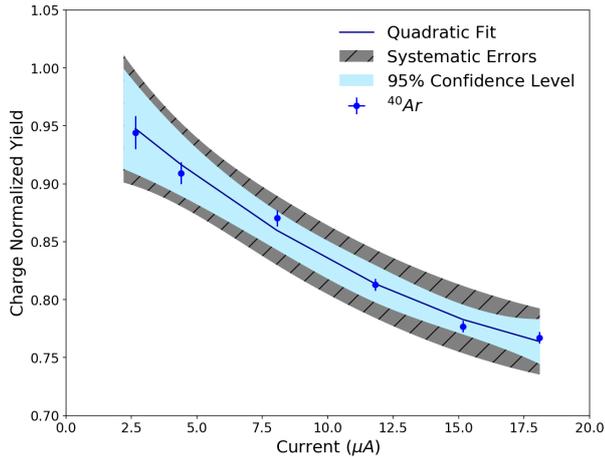}
	\caption{Shown is the $^{40}$Ar target's local density as a function of beam current.}
	\label{fig:argon_data}
\end{figure}

\begin{figure*}[htb]
\begin{center}
  \begin{subfigure}{0.45\textwidth}
    \centering\includegraphics[width=\textwidth]{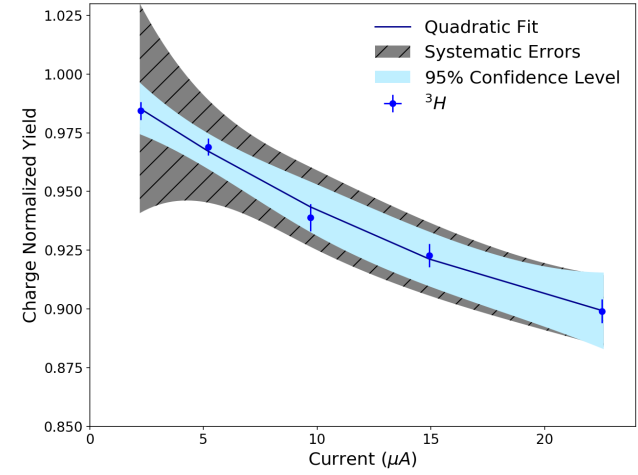}
    \caption{$^{3}$H Density Analysis. }
    \label{fig:tritium_data}
  \end{subfigure}
  \begin{subfigure}{0.45\textwidth}
    \centering\includegraphics[width=\textwidth]{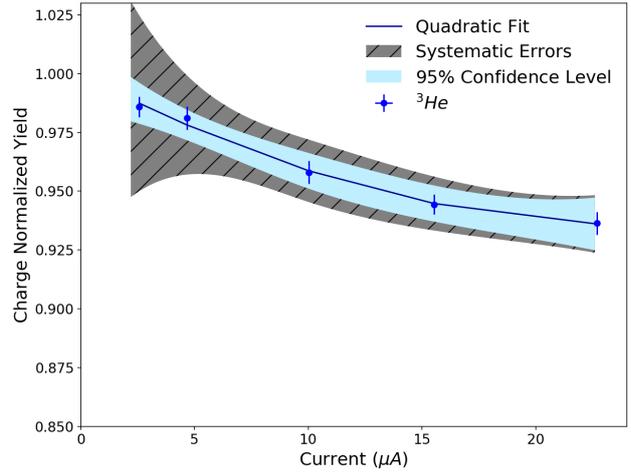}
    \caption{$^{3}$He Density Analysis.}
    \label{fig:helium_data}
  \end{subfigure}
  \begin{subfigure}{0.45\textwidth}
    \centering\includegraphics[width=\textwidth]{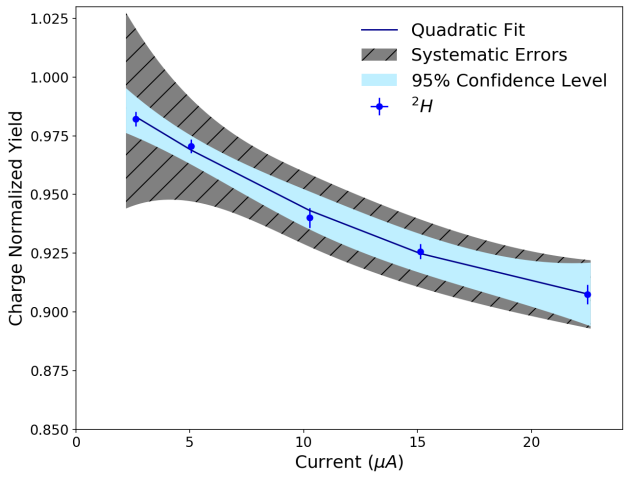}
    \caption{$^{2}$H Density Analysis.}
    \label{fig:deuterium_data}
  \end{subfigure}
  \quad
  \begin{subfigure}{0.45\textwidth}
    \centering\includegraphics[width=\textwidth]{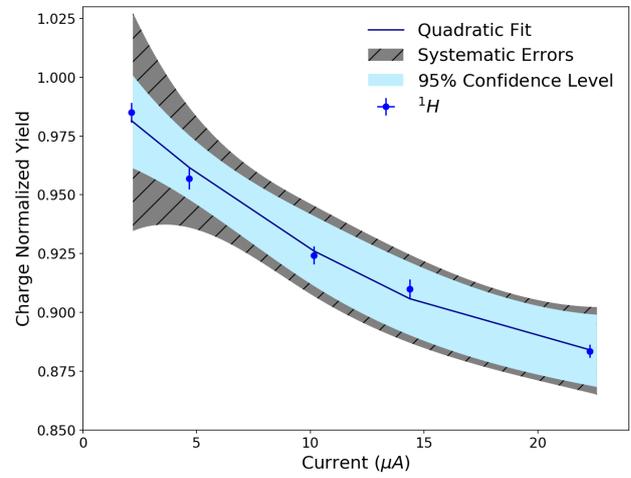}
    \caption{$^{1}$H Density Analysis.}
    \label{fig:hydrogen_data}
  \end{subfigure}
  \end{center}
  \label{fig:tritium_targets}
  \caption{Shown is local density of the $^{3}$H, $^{3}$He, $^{2}$H and $^{1}$H targets
           as a function of beam current.}
\end{figure*}

\begin{figure}[htb]
 \centering
 \includegraphics[width=0.9\linewidth]{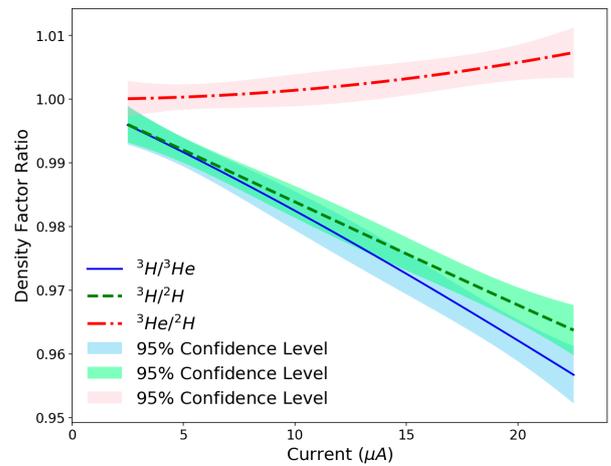}
  \caption{For experiments that will be taking the ratios between different targets, we also determined
          the ratio the density changes.   As some of the systematic affects cancel in the ratio, these uncertainties are 
          slightly smaller then the absolute density change determinations.}
\label{fig:density_ratios}
\end{figure}

The density correction was determined for each gas species by measuring $Y_{\rm{norm}}$ as a function of 
beam current $I_{\rm{beam}}$. The function is then normalized to $1$ for $I_{\rm{beam}}=0$. The density 
each gas cell for zero beam current is the same as that of the load density.  Figures~\ref{fig:argon_data}, 
\ref{fig:tritium_data}, \ref{fig:helium_data}, \ref{fig:deuterium_data} and \ref{fig:hydrogen_data} 
show the density correction for the different gas targets. It is easily seen that the density decreases 
with the current and that the behavior of the density correction factor $f$ is modeled well 
by a quadratic function

\begin{equation}
f(I_{\rm{beam}}) = a\cdot I_{\rm{beam}}^{2} + b \cdot I_{\rm{beam}} + c.
\label{eq:boiling_factor}
\end{equation}

\noindent where $a$, $b$ and $c$ are the fit parameters. Table~\ref{tab:fit_parameters} shows the 
fit parameters for each gas species. The density correction factor $f(I_{\rm{beam}})$ 
is determined for each gas by substitution of these parameters in Eq.~\ref{eq:boiling_factor}. 
The density correction factor determined in this manner is valid for the current range $0-22.5$ $\mu A$. 
The error bar in the plots represents the statistical uncertainty only, and a fit was calculated 
with respect to those values with a $95\%$ confidence band in blue. The gray hatched $95\%$ 
confidence band represents a fit including both statistical and systematic uncertainties.
Since many data analyses require ratios of different targets,
we also provide the ratio of the density changes, shown in Fig.~\ref{fig:density_ratios}.

\begin{table}[htb]
\begin{tabular}{|c|c|l|c|c|l|}
\hline
\multicolumn{3}{|c|}{\textbf{$^{3}$H Fit Parameters}}                                & \multicolumn{3}{l|}{\textbf{$^{3}$H Correlation Factors}}    \\ \hline
\textbf{a}              & \multicolumn{2}{c|}{$(1.06 \pm 0.36) \times 10^{-4}$}                & \textbf{C(a, b)}             & \multicolumn{2}{c|}{$-0.974$} \\ \hline
\textbf{b}              & \multicolumn{2}{c|}{$(-6.8 \pm 0.89) \times 10^{-3}$}                 & \textbf{C(b, c)}             & \multicolumn{2}{c|}{$-0.888$} \\ \hline
\textbf{c}              & \multicolumn{2}{c|}{$1. +/- 0.003$}                        & \textbf{C(a, c)}             & \multicolumn{2}{c|}{$0.801$}  \\ \hline
\multicolumn{3}{|c|}{\textbf{$^{3}$He Fit Parameters}}                               & \multicolumn{3}{c|}{\textbf{$^{3}$He Correlation Factors}}   \\ \hline
\textbf{a}              & \multicolumn{2}{c|}{$(1.04 \pm 0.25) \times 10^{-4}$}                 & \textbf{C(a, b)}                      & \multicolumn{2}{c|}{$-0.973$} \\ \hline
\textbf{b}              & \multicolumn{2}{c|}{$(-5.1 \pm 0.64) \times 10^{-3}$}                 & \textbf{C(b, c)}                      & \multicolumn{2}{c|}{$-0.879$} \\ \hline
\textbf{c}              & \multicolumn{2}{c|}{$1 \pm 0.003$}                         & \textbf{C(a, c)}                     & \multicolumn{2}{c|}{$0.779$}  \\ \hline
\multicolumn{3}{|c|}{\textbf{$^{2}$H Fit Parameters}}                                & \multicolumn{3}{c|}{\textbf{$^{2}$H Correlation Factors}}    \\ \hline
\textbf{a}              & \multicolumn{2}{c|}{$(1.16 \pm 0.29) \times 10^{-4}$} & \textbf{C(a, b)}             & \multicolumn{2}{c|}{$-0.973$} \\ \hline
\textbf{b}              & \multicolumn{2}{c|}{$(-6.7 \pm 0.71) \times 10^{-3}$}                 & \textbf{C(b, c)}             & \multicolumn{2}{c|}{$-0.895$} \\ \hline
\textbf{c}              & \multicolumn{2}{c|}{$1. \pm 0.003$}                        & \textbf{C(a, c)}             & \multicolumn{2}{c|}{$0.805$}  \\ \hline
\multicolumn{3}{|c|}{\textbf{$^{1}$H Fit Parameters}}                                & \multicolumn{3}{c|}{\textbf{$^{1}$H Correlation Factors}}    \\ \hline
\textbf{a}              & \multicolumn{2}{c|}{$(1.70 \pm 0.47) \times 10^{-4}$ }               & \textbf{C(a, b)}             & \multicolumn{2}{c|}{$-0.978$} \\ \hline
\textbf{b}              & \multicolumn{2}{c|}{$(-9 \pm 0.12) \times 10^{-3}$   }                & \textbf{C(b, c)}             & \multicolumn{2}{c|}{$-0.881$} \\ \hline
\textbf{c}              & \multicolumn{2}{c|}{$1. \pm 0.006$}                        & \textbf{C(a, c)}             & \multicolumn{2}{c|}{$0.788$}  \\ \hline
\multicolumn{3}{|c|}{\textbf{$^{40}$Ar Fit Parameters}}                              & \multicolumn{3}{c|}{\textbf{$^{40}$Ar Correlation Factors}}  \\ \hline
\textbf{a}              & \multicolumn{2}{c|}{$(4.33 \pm 1.5) \times 10^{-4}$}               & \textbf{C(a, b)}             & \multicolumn{2}{c|}{$-0.981$} \\ \hline
\textbf{b}              & \multicolumn{2}{c|}{$(-2.1 \pm 0.3) \times 10^{-2}$}                    & \textbf{C(b, c)}             & \multicolumn{2}{c|}{$-0.942$} \\ \hline
\textbf{c}              & \multicolumn{2}{c|}{$1. \pm 0.02$}                         & \textbf{C(a, c)}             & \multicolumn{2}{c|}{$0.867$}  \\ \hline
\end{tabular}
\caption{Fit parameters obtained for the percentage of density change calculation with respect to the beam current.}
\label{tab:fit_parameters}
\end{table}

\subsection{Systematic Uncertainties}

Several corrections are applied to the data in this analysis, and since the current is different 
for every point, the uncertainties are evaluated at every point.   Confidence bands for each fit 
including the systematic uncertainties are shown in Fig.~\ref{fig:argon_data}, \ref{fig:tritium_data}, 
\ref{fig:helium_data}, \ref{fig:deuterium_data} and \ref{fig:hydrogen_data}.  They include the 
uncertainty in the charge, live-time and detector efficiencies.

The BCM monitors are effective over a range from 0 to 100~$\mu$A. However, low current measurements 
have a slightly higher uncertainty causing the uncertainty in the charge to be current dependent. 
The uncertainty in the current and charge is estimated using the BCM calibration shown in Fig.~\ref{fig:dnew_cal}, 
together with the error covariance matrix.  This is the dominant source of systematic uncertainty in 
the determination of the density reduction factor $f(I_{\rm{beam}})$.

The background contamination coming from the entrance and exit windows is also a source 
of systematic uncertainty. This is due to the thickness variations in the cell entrance and 
exit windows which can be seen in Figure \ref{fig:bk_empty}. Therefore, in order to calculate 
the background uncertainty in the measurement, the percentage of background was calculated 
in $y_{\rm{tar}}$ for the values of $\pm$4~cm, $\pm$7~cm and 
$\pm$ 10~cm from the center of the target.  The same normalization procedure was followed for each of the 
different cuts in the reaction vertex region to calculate $f(I_{\rm{beam}})$. Finally, 
the uncertainty in the background contamination is given by the standard deviation 
of the average of multiple $f(I_{\rm{beam}})$ obtained with the different cuts. 
The standard deviation was never more than 1$\%$ for each current.

Furthermore, $1\%$ systematic uncertainties were estimated for the live-time, VDC one-track efficiency, 
trigger efficiency, detector and cut efficiencies of the gas Cherenkov and $\pi^{-}$ rejection.

\section{Summary}

A novel design for low density gas targets has been used in the Jefferson Lab electron beam 
with a number of different gas species.  These cells have proven to be extremely robust and 
satisfied the safety requirements necessary for holding 1~kCi of tritium gas.
In this manuscript, we have shown how the local density of these cells changes when an electron beam
passes through them.  The 5\% to 10\% changes that were measured at 22.5~uA for the different 
gas species are consistent with the design expectations.    Determination these density changes
was critical for experiments using these cells for cross section measurements.

\section{Acknowledgments}

We wish to thank the staff of the Thomas Jefferson National Accelerator Facility for their
help safely installing and removing the tritium cells.
We also acknowledge the critical efforts of Savannah River Site and
Savannah River Tritium Enterprises.
And special thank to Marcy Stutzman for proofreading this manuscript.
This work was supported by the U.S. Department of Energy (DOE) contract
DE-AC05-06OR23177 under which Jefferson Science Associates operates 
the Thomas Jefferson National Accelerator Facility,
DOE contract DE-AC02-06CH11357, DOE contract DE-SC0013615, 
and by National Science Foundation (NSF) Grant No. NSF PHY 1506459.

\section*{References}
\bibliography{references}

\end{document}